\documentclass[prl,twocolumn,amsmath,amssymb,aps,preprintnumbers,superscriptaddress]{revtex4-1}
\usepackage{physics}
\usepackage[normalem]{ulem}
\usepackage{bbold}
\usepackage{braket}
\usepackage{amsmath, amsfonts, amssymb}
\usepackage{graphicx}
\usepackage{float}
\usepackage{stackengine}
\usepackage[colorlinks=true,linkcolor=blue,citecolor=blue,allcolors=blue]{hyperref}
\usepackage[caption=false]{subfig}
\usepackage{mathtools}
\usepackage{tikz}
\usetikzlibrary{tikzmark}
\usepackage{cancel}

\usepackage{notes2bib}
\usepackage{graphicx}
\bibnotesetup{
	note-name = ,
	use-sort-key = false
}

\usepackage{lipsum}
\usepackage{setspace}
\newcommand{\nobarfrac}{\genfrac{}{}{0pt}{}}
\begin{document}
	\newcommand{\titleinfo}{Formation of Rydberg Crystals Induced by Quantum Melting in One-Dimension}
	\title{\titleinfo}
	
	\author{Zeki Zeybek}
	\email{zzeybek@physnet.uni-hamburg.de}
	\affiliation{The Hamburg Centre for Ultrafast Imaging, Universit{\"a}t Hamburg, Luruper Chaussee 149, 22761 Hamburg, Germany}
	\affiliation{Zentrum f{\"u}r Optische Quantentechnologien, Universit{\"a}t Hamburg, Luruper Chaussee 149, 22761 Hamburg, Germany}

	\author{Peter Schmelcher}
	\affiliation{The Hamburg Centre for Ultrafast Imaging, Universit{\"a}t Hamburg, Luruper Chaussee 149, 22761 Hamburg, Germany}
	\affiliation{Zentrum f{\"u}r Optische Quantentechnologien, Universit{\"a}t Hamburg, Luruper Chaussee 149, 22761 Hamburg, Germany}
	
	\author{Rick Mukherjee}
	\email{rmukherj@physnet.uni-hamburg.de}
	\affiliation{Zentrum f{\"u}r Optische Quantentechnologien, Universit{\"a}t Hamburg, Luruper Chaussee 149, 22761 Hamburg, Germany}
	\begin{abstract}
 Quantum fluctuations in frustrated systems can lead to the emergence of complex many-body phases. However, the role of quantum fluctuations in frustration-free lattices is less explored and could provide an interesting avenue for exploring new physics, and perhaps easier to realize compared to frustrated lattice systems. Using Rydberg atoms with tunable interactions as a platform, we leverage strong van der Waals interactions and obtain a constrained model in one dimension with non-local fluctuations given by dipolar interactions alongside local fluctuations. The combined effect of such processes leads to intrinsically quantum-ordered Rydberg crystals through the order-by-disorder mechanism. Finite-size analyses indicate that combined fluctuations drive the transition from disordered to ordered phases, contrary to the expected direction. We provide a theoretical description to understand the physics of order-by-disorder in one-dimensional systems, which are typically seen only in higher dimensions.
	\end{abstract}
	\maketitle
        
        \textit{Introduction.\textemdash}Quantum fluctuations give rise to diverse phenomena ranging from spin fluctuations induced superfluidity of helium-3 \cite{SF_Anderson_1973, SF_Anderson_1974} to superconductivity in metals \cite{SC_Appel_1980}. In the case of frustrated systems, quantum fluctuations lift the macroscopic classical ground-state degeneracy \cite{millonas_fluctuations_1996, ann_rev_QOBD_2018, RevMod_QSpinLiq_2017, balents_spin_2010, intro_frust_mila} and stabilize exotic many-body phases such as the quantum spin liquids \cite{RevMod_QSpinLiq_2017, balents_spin_2010}, quantum spin nematics \cite{spinNem1, spinNem2, mulder_ObD_2010}, long-range ordered states through order-by-disorder without any classical analog \cite{villain_first_1980, ObD_Henley, ObD_Henley2, Obd_Honeycomb_J1J3_FerroAFMod, sachdev_QObD_Kagome_1992, starykh_unusual_2015, rev_pyrochlore}, as well as lead to complex phase transitions \cite{ObD_Emergent_PottsOrder_Kagome_J1_J3_Heisenberg, ObD_Emerg_Potts_NematicSLiq_Kagome_J1_J3_Heisenberg, ObD_PseudoGold_ThermalObD}. However, it can often be hard to realize these phases in frustrated materials that involve non-trivial geometries such as kagome \cite{Quantum_Thermal_ObD_KagoemAF_Chiral_Int, ObD_Emergent_PottsOrder_Kagome_J1_J3_Heisenberg, ObD_Emerg_Potts_NematicSLiq_Kagome_J1_J3_Heisenberg}, honeycomb \cite{ObD_KitaevMagnet, Obd_Honeycomb_J1J3_FerroAFMod, ObD_SpinOrbitalLiquids}, triangle \cite{ObD_Triangle_Hubbard,Quantum_Thermal_ObD_Triangle_SU3Heisenberg_w_BField}, and pyrochlore \cite{ObD_Experimental_XY_Pyrochlore, XY_pyrochhlore} lattices. Further issues arise from environment noise, inherent disorder in the system and the limited control of competing interactions \cite{rev_pyrochlore,Experimental_Problems_Pyrochlore}. 
        
        Platforms based on neutral Rydberg atoms have proven to be highly controllable quantum simulators \cite{QSimRyd_1, QSimRyd_2, QsimRyd_3, QSimRyd_4, ebadi_quantum_2021, scholl_quantum_2021} as their large dipole moments provide tunable strong interactions with different ranges and characters such as van der Waals (vdW) and dipolar interactions \cite{weimer_rydberg_2010, browaeys_experimental_2016, browaeys_many-body_2020, whitlock_simulating_2017}. For example, the quantum spin liquid phase was observed in a recent study involving vdW interacting Rydberg atoms \cite{verr, semeghini_probing_2021}. However, fluctuation-driven phenomena were not explored using both vdW and dipolar interactions, and there is a growing interest in studying the combined effects of these processes \cite{zeybek_2023,zeybek2024bondorder,de_leseleuc_observation_2019, scholl_microwave_2022, weidmull}. Especially, a question arises on whether the flexibility of tuning both vdW and dipolar interactions can be leveraged to observe such phenomena in simple frustration-free setups such as a one-dimensional (1D) lattice. 
        
        In this Letter, we address this question by studying a 1D system of vdW and dipolar interacting Rydberg atoms and reveal the relationship between the system geometry and competing fluctuations. In the limit of strong vdW interactions, a constrained model where local and non-local quantum fluctuations are driven by single- and two-site processes is derived. Having such different melting processes simultaneously leads to a phase with long-range order via the order-by-disorder mechanism, which has not been shown previously in frustration-free systems in 1D. We identify the phase as a form of Rydberg crystal without any classical analog since it does not arise by minimizing the vdW energy, differing from earlier results on Rydberg crystals \cite{weimer_two-stage_2010,sela_dislocation-mediated_2011,bernien_probing_2017}. We present an intuitive physical picture of the underlying physics by providing a theoretical analysis of the model. This helps explain the long-range order and motivates a variational ansatz that describes both the disordered and ordered regimes. The nature of the quantum phase transition (QPT) is examined by finite-size scaling analyses. We reveal that the combination of local and non-local quantum fluctuations drives a phase transition that proceeds in the disordered-to-ordered direction, which conventionally occurs the other way around.

\begin{figure}[t!]
		\includegraphics[width=\columnwidth]{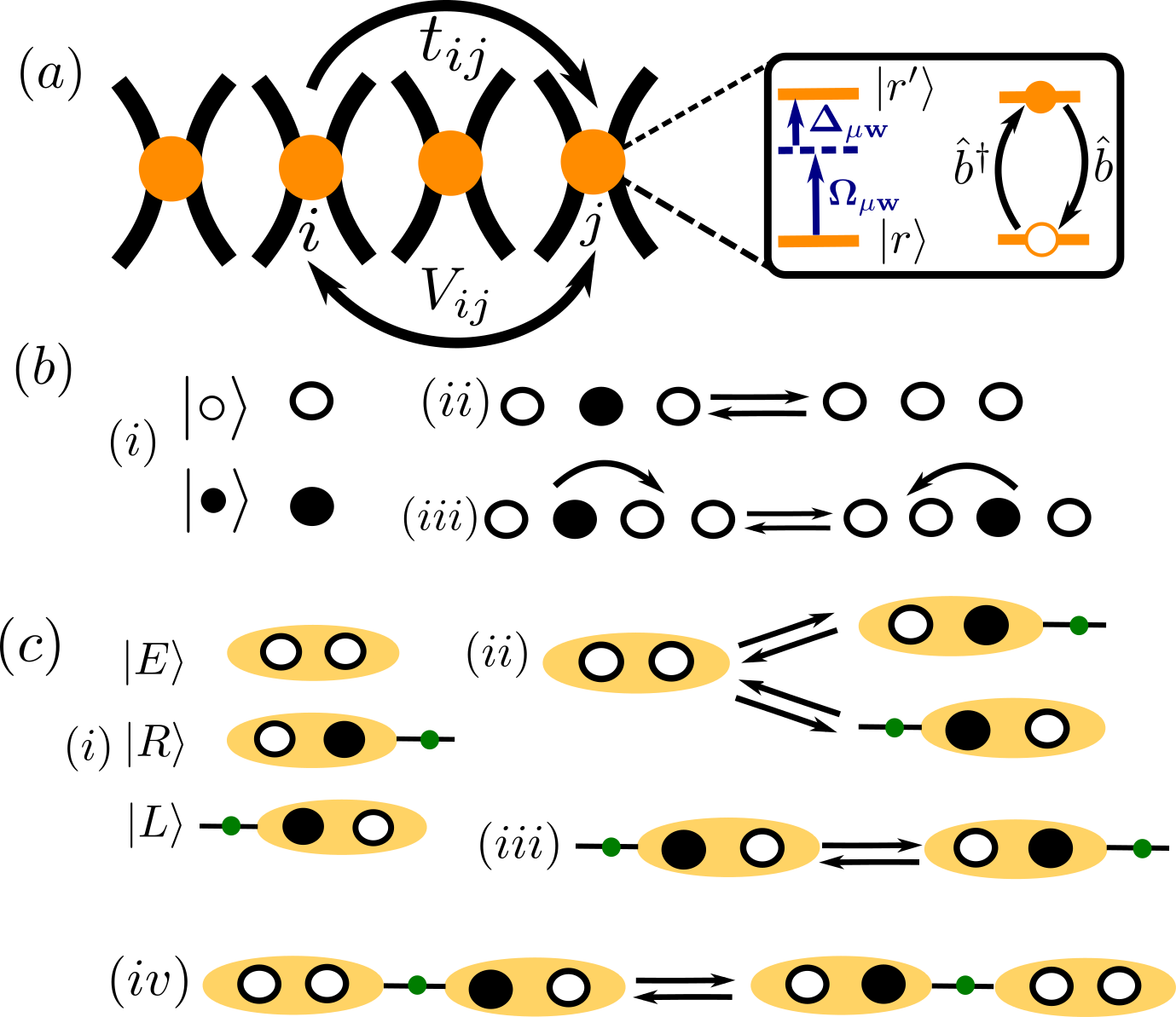}
		\caption{(a) Schematic of a 1D system of vdW $(V_{ij})$ and dipolar $(t_{ij})$ interacting Rydberg atoms. Coupling of the levels $\ket{r} \leftrightarrow \ket{r^{\prime}}$ by a microwave laser with Rabi frequency $\Omega_{\mu w}$ and detuning $\Delta_{\mu w}$ corresponds to the creation (annihilation) of a boson with $\hat b^\dagger (\hat b)$. (b.i) Single-site degrees of freedom are denoted by the empty $\ket{\circ} \rightarrow \circ$ and boson-occupied sites $\ket{\bullet} \rightarrow \bullet$. (b.ii) Constrained local fluctuations generated by single-site processes as creation/annihilation of bosons surrounded by empty sites. (b.iii) Constrained non-local fluctuations provided by two-site processes as hopping of bosons surrounded by empty sites. (c.i) Dimer degrees of freedom are given by combining two sites into one, which makes up a single dimer site. Auxiliary fermions are attached to links connecting dimer sites to ensure the occupation constraint (see the text). Right dimer $\ket{R}$ and left dimer $\ket{L}$ states are given by right and left occupied two-sites with fermions attached (green dot) to the right and left links (black line). Processes involving (c.ii) left/right dimer creation/annihilation and (c.iii) dimer flipping at a given dimer site are depicted. (c.iv) Dimer exchange between different dimer sites is shown.} 
		\label{fig:sett}
\end{figure}

 \textit{Theory.\textemdash}We consider a 1D system of trapped neutral atoms that are modeled as two-level systems with a pair of Rydberg states coupled by a microwave laser with the Rabi frequency $\Omega_{\mu w}$ and detuning $\Delta_{\mu w}$ as shown in Fig.~\ref{fig:sett}(a). These distinct highly excited Rydberg states represent the hard-core bosonic degree of freedom given by $\{ \ket{\circ},\ket{\bullet}\}$ where $\ket{\circ}$ ($\ket{\bullet}$) denotes the absence (presence) of a boson [Fig.~\ref{fig:sett}(b.i)]. This encoding leads to expressing the vdW and dipolar interacting Rydberg atoms by an extended Bose-Hubbard model (EBHM) \cite{de_leseleuc_observation_2019,zeybek_2023, zeybek2024bondorder} as given in the following,
         \begin{align}\label{eqn:EBHM_Ryd} 
		\hat H_{\text{Ryd}} &= \sum_{i<j} t_{ij}( \hat b^{\dagger}_i \hat b_j +\text{h.c.}) + \sum_{i<j}V_{ij} \hat n_i \hat n_j \notag  -  \Delta_{\mu w}\sum_{i}\hat n_i \\&+\Omega_{\mu w}\sum_{i}(\hat b^{\dagger}_i+ \hat b_i) ,
	\end{align}
where $\hat b^{\dagger}_i$ ($\hat b_i$) is the bosonic creation (annihilation) operator at site $i$ with $(\hat b^{\dagger}_i)^2=0$, and $\hat n_i=\hat b^{\dagger}_i \hat b_i$ is the number operator. vdW interactions give rise to long-range density interaction with strength $V_{ij}>0$ with $1/|i-j|^6$ scaling. The detuning $\Delta_{\mu w}$ term corresponds to the chemical potential which controls the density of excitations $\ket{r^{\prime}}$ (bosons). Differing from other EBHMs, Eq.~(\ref{eqn:EBHM_Ryd}) includes both local and non-local quantum fluctuations that are provided by the Rabi term and the dipolar interactions involving singe-site and two-site processes, respectively. The Rabi term $\Omega_{\mu w}$ introduces particle fluctuations at a site with creation/annihilation processes. The dipolar interactions encode particle exchange involving two sites with strength $t_{ij}>0$ with $1/|i-j|^3$ scaling. Such single- and two-site processes lead to the local and non-local melting of crystalline phases into a disordered and Luttinger liquid phase \cite{zeybek_2023}, respectively. 

In the classical limit ($\Omega_{\mu w},t \rightarrow 0$) of Eq.~(\ref{eqn:EBHM_Ryd}) with only nearest-neighbor interactions and $\Delta_{\mu w}=0$, all the configurations without nearest-neighbor boson occupation are degenerate with energy $E = 0$ and the degeneracy scales exponentially as $e^{L \ln \alpha}$ with the system size $L$ and $\alpha=(\sqrt{5}+1)/2$ \cite{Deg_size_Domb_1960, Deg_size_Bax_1960}. This can form the required degenerate classical ground-state manifold in the limit of strong nearest-neighbor interactions ($V_{i,i+1} \rightarrow \infty $) where the Hilbert space is constrained such that configurations with boson occupation in nearest-neighbor sites such as $\ket{\dots \bullet \bullet \dots}$ are excluded. Therefore, the degenerate manifold is provided by the occupation constraint, differing from the models with frustration. The resulting model in this projected subspace is obtained perturbatively \bibnote[SM]{See Supplemental Material for the derivation of the constrained model, details on the dimer description and the variational ansatz with additional numerical results, which includes Refs. \cite{QPT_Book_Sachdev_2011, SWT_Cohen, SWT_REV_2011,ChengI, Cheng_II, ChengIII}}. With vanishing longer-range terms ($t_{i,i+k}, V_{i,i+k} = 0, k>1$) and detuning ($\Delta_{\mu w}  = 0$), the Hamiltonian up to first order is expressed as,

 \begin{align}\label{eqn:PXP}
		\hat H &= \Omega_{\mu w}\sum_i \hat P_{i-1}(\hat b^\dagger_i +\hat b_i) \nonumber  \hat P_{i+1}\\&+ t\sum_{i}\hat P_{i-1}(\hat b^\dagger_i \hat b_{i+1} + \text{h.c.})\hat P_{i+2},
\end{align}
where $\hat P_i = \ket{\circ}_i \bra{\circ}$ is the projector operator to the absence of a boson. In the first term, dressing the local fluctuations with the projectors leads to the well-studied PXP Hamiltonian \cite{PXP_FSS, turner_2018, turner_weak_2018, PXP_REF1, PXP_REF2, bluvstein_PXP_ETH}. It involves single-site processes [Fig.~\ref{fig:sett}(b.ii)] where creation/annihilation at a given site is allowed if empty sites surround it. Configurations with different particle numbers are coupled, therefore, $U(1)$-symmetry is violated. In the second term, dressing the non-local fluctuations [Fig.~\ref{fig:sett}(b.iii)] with the projectors gives rise to constrained hoppings \cite{alcaraz1999exactly,verresen2019stable} where empty sites must surround the two sites involved in the particle exchange. Such processes preserve $U(1)$-symmetry since the particle number is conserved. The interplay between local and non-local fluctuations dictates the ground state properties of the model. Thus, Eq.(\ref{eqn:PXP}) essentially describes the onset of the competing local and non-local fluctuations for determining which states from the degenerate manifold are to be favored. Differing from previous models, this aspect will play a crucial part in stabilizing long-range order in 1D, which has been previously reported to exist only in higher dimensional constrained/frustrated models \cite{yue_bhaskar_2021,sarkar_2023}.

We consider a theoretical description of the model where we split the lattice into interacting dimers consisting of two sites. This artificial partitioning into dimers will prove to be important in explaining the physics of the disordered and ordered regions. With this partitioning, a single dimer site $j$ consists of two sites as $j \equiv (i,i+1)$ and encodes the dimer degrees of freedom with the allowed configurations given by $\{\ket{\circ \circ},\ket{\circ \bullet},\ket{\bullet \circ}\}$.  We attach a fermion to the left (right) link of the configuration $\ket{\bullet \circ}$ ($\ket{\circ \bullet}$) which we denote as left(right)-dimer $\ket{L}$ ($\ket{R}$), and the configuration $\ket{\circ\circ}$ corresponds to the empty dimer $\ket{E}$ [Fig.~\ref{fig:sett}(c.i)]. Fermionic links naturally implement the occupation constraint where the configurations of the form $\ket{\dots RL \dots}$ are excluded due to the Pauli exclusion principle. Representing the system in terms of dimer degrees of freedom can be considered as distinguishing even and odd site boson occupations in the form of a mapping $\hat{P}_{i-1}\hat b^{\dagger}_{i}\hat P_{i+1} \to \hat d^{\dagger}_{R,j} \hat f^{\dagger}_{j,j+1} $ with $i=2j$ and $\hat{P}_{i-1}\hat b^{\dagger}_{i}\hat P_{i+1} \to d^{\dagger}_{L,j} \hat f^{\dagger}_{j-1,j}$ with $i=2j-1$. This mapping is discussed in detail in \cite{SM}. In terms of dimer states, the Hamiltonian~(\ref{eqn:PXP}) is rewritten as,
\begin{align}\label{eq: DIM}
    \hat H_{\text{dim}} &= \Omega_{\mu w}\sum_j ( \hat d^{\dagger}_{L,j}\hat f^{\dagger}_{j-1,j} + \hat d^{\dagger}_{R,j}\hat f^{\dagger}_{j,j+1} \nonumber + \text{h.c.}) \\ &+  t\sum_j (\hat d^{\dagger}_{L,j}\hat f^{\dagger}_{j-1,j}\hat d_{R,j}\hat f_{j,j+1}+ \text{h.c.} ) \nonumber \\
    &+ t\sum_{\expval{j,k}}(\hat d^{\dagger}_{R,j}\hat f^{\dagger}_{j,j+1}\hat d_{L,k}\hat f_{k,k-1}+ \text{h.c.}),
\end{align}
where left $\hat d^{\dagger}_{L,j}$ ($\hat d_{L,j}$) and right $\hat d^{\dagger}_{R,j}$ ($\hat d_{R,j}$) are dimer creation (annihilation) operators at a dimer site $j$ and $\hat f^{\dagger}_{j,j+1}$ ($\hat f_{j,j+1}$) creates (annihilates) fermions at the links between dimer sites ($j,j+1$). The first term above is the creation/annihilation of left and right dimers [Fig.~\ref{fig:sett}(c.ii)]. The second term encodes a local dimer flipping where $\ket{L}_j$ is coupled to $\ket{R}_j$ [Fig.~\ref{fig:sett}(c.iii)]. The third term corresponds to a dimer exchange between nearest-neighbor sites $\expval{j,k}$ where $\ket{L}_j$ is exchanged with $\ket{R}_k$ [Fig.~\ref{fig:sett}(c.iv)]. Motivated by this formulation, we propose an ansatz to describe the ground state in the following form, 
\begin{align}
    \ket{\Psi} = \frac{1}{\mathcal{N}}\prod_j (u+ v \hat d^{\dagger}_{L,j}\hat f^{\dagger}_{j-1,j}+ w \hat d^{\dagger}_{R,j}\hat f^{\dagger}_{j,j+1})\ket{E \cdots E}
\end{align}
which can be written in terms of dimer states as

\begin{equation}\label{eq:ansatz}
     \ket{\Psi} = \frac{1}{\mathcal{N}}\sum_{\Lambda} u^{N^{\Lambda}_{E}}  v^{N^{\Lambda}_{L}} w^{N^{\Lambda}_{R}} \ket{\Lambda},
\end{equation}
where $N^{\Lambda}_{E,L,R}$ denotes the number of empty, left and right dimers in the many-body configuration $\ket{\Lambda}$ of dimer states. The variational parameters are $u,v,w$ and $\mathcal{N}$ is the normalization constant. The chosen ansatz is well-suited to capture the ground state properties as it addresses various aspects of the system: $i)$ Configurations $\ket{\Lambda}$ automatically satisfy the occupation constraint. $ii)$ The coherent superposition form of the ansatz can resonate a subset of classical configurations $\ket{\Lambda}$ which may be favored by the quantum fluctuations. $iii)$ The superposition can include configurations with different numbers of left/right dimers. Therefore, the ansatz can interpolate between $U(1)$-preserving/violating regions accounting for the effects of both local and non-local fluctuations. $iv)$ The weights of the configurations $\ket{\Lambda}$ in the superposition are determined according to the number of dimer types therein. Therefore, the correlations emanating from the fluctuations of the dimer densities in a given $\ket{\Lambda}$ can be captured. 

\textit{Results.\textemdash} We demonstrate that even in a simple 1D lattice, the combination of local- and non-local quantum fluctuations helps stabilize long-range order, even though they do not promote order individually. The order formation is elucidated using the theoretical description involving interacting dimers and the variational ansatz. The nature of the QPT is examined by performing finite-size-scaling analyses. Numerical results are obtained by employing density-matrix-renormalization-group \cite{hauschild_efficient_2018,white_density_1992,white_density-matrix_1993,scholwork2005,SCHOLLWOCK2011} simulations.

\begin{figure}[!t]
		\includegraphics[width=1\columnwidth]{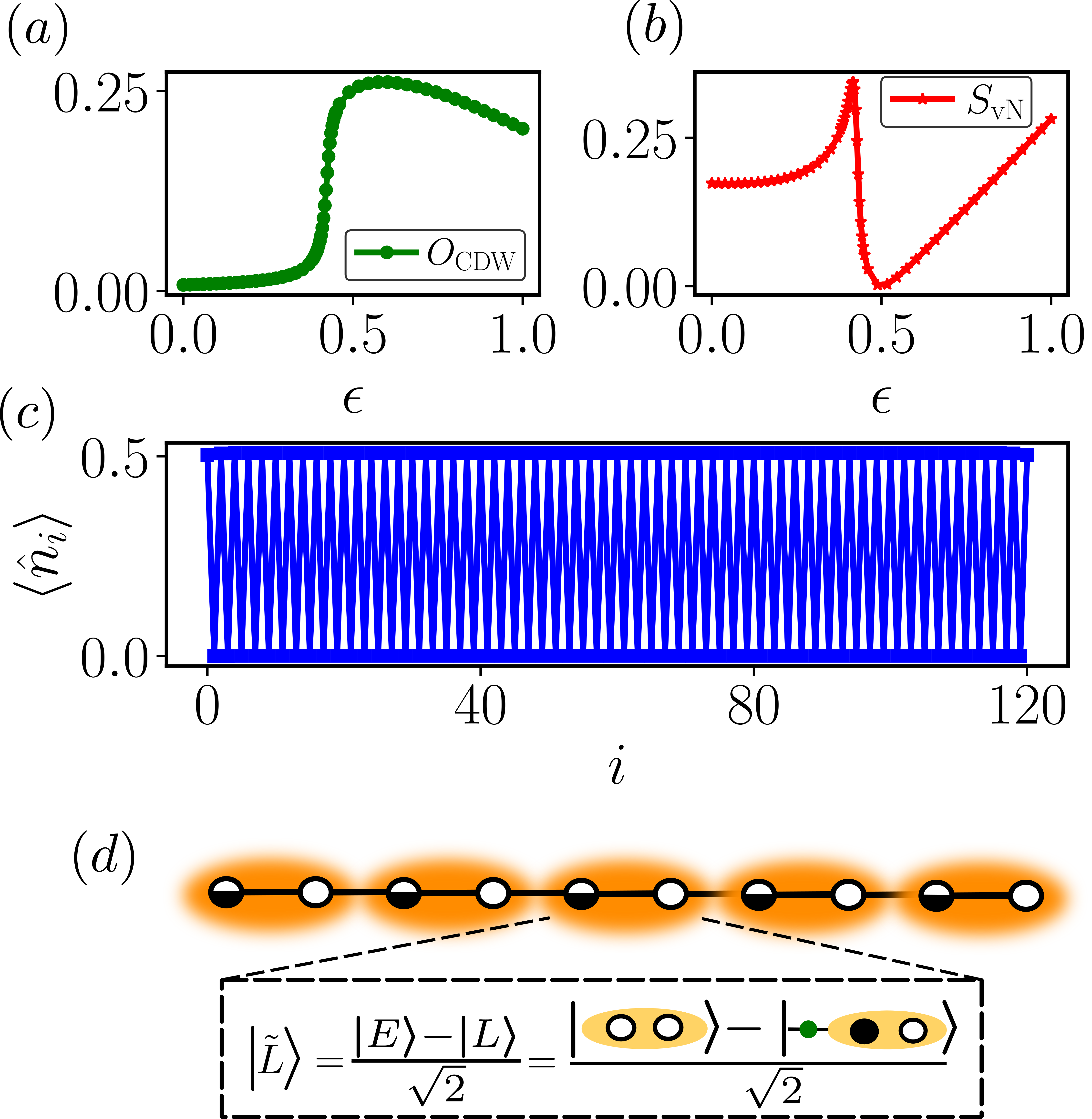}
		\caption{(a) CDW order parameter $O_{\text{CDW}}$ and the (b) bipartite von Neumann entanglement entropy $S_{\text{vN}}$ as a function of $\epsilon=t/\Omega_{\mu w}$ is depicted. (c) The expectation value of the density operator $\hat n_i$ for a system size $L=121$ with $\epsilon=0.5$ is shown. (d) The schematic of the ground state with CDW order is displayed. The unit cell is given by the equal superposition of $\ket{E}$ and $\ket{L}$ as $\ket{\tilde L} = (\ket{E}-\ket{L})/\sqrt{2}$.} 
		\label{fig:ord_par}
\end{figure} 

We start discussing two different regimes of $\epsilon=t/\Omega_{\mu w}$ with the charge-density-wave (CDW) order parameter $O_{\text{CDW}}$ and entanglement entropy $S_{\text{vN}}$. In the limit $\epsilon \rightarrow 0$, local quantum fluctuations become dominant and lift the degeneracy by resonating classical configurations. In this mechanism, the classical configurations from the degenerate manifold are selected such that their superposition state can have low-energy fluctuations when acted upon by the $\Omega_{\mu w}$ term. For example, the configuration $\ket{\bullet \circ \bullet \cdots \bullet \circ}$ is not favored since $\Omega_{\mu w}$ term creates high-energy fluctuations in the form $\ket{\bullet \bullet \bullet \cdots \bullet \circ}$ by violating the occupation constraint. Therefore, the classical configurations with a high density of low-energy mode are favored when acted upon by the $\Omega_{\mu w}$ term. In 1D, this stabilizes a disordered phase exhibiting homogeneous density without long-range order. This can be seen in Fig.~\ref{fig:ord_par}(a) with vanishing $O_{\text{CDW}}= (1/L)\sum_i (-1)^{i}\expval{\hat n_i}$ which measures the staggered density. To gain more insight into the superposition, we compute the overlap of the ansatz $\ket{\Psi}$ in Eq.~(\ref{eq:ansatz}) with the exact ground state $\ket{\Psi_{\text{ED}}}$ and obtain $|\braket{\Psi|\Psi_{\text{ED}}}| \sim 0.99$ with the variational parameters $u=0.92,v=-0.58,w=-0.58$ \cite{SM}. Equal weight given to the left/right dimer state demonstrates no preference for selecting either side. We find that the optimized ansatz $\ket{\Psi}$ can be composed by $\bigotimes_i \ket{\mathcal{D}}_i$ with $\ket{\mathcal{D}}_i=[(\ket{E}_i/\sqrt{2})-(\ket{L}_i+\ket{R}_i) /2]$ and $i$ denoting a single dimer site \cite{SM}. This implies that each dimer site has an equal probability of being in $\ket{L}$ or $\ket{R}$. In light of this, the role of the geometry becomes apparent when the state selection mechanism from the degenerate manifold is considered. Due to the low connectivity in 1D, both the left and right dimer dominant $\ket{\Lambda}$ configurations have low-energy fluctuations due to the $\Omega_{\mu w}$ term, thereby, the disordered phase with equal favoring of $\ket{L}$ and $\ket{R}$ is promoted. Compared to 2D lattices where order-by-disorder has been extensively studied, configurations that equally favor $\ket{L}$ and $\ket{R}$ are not seen as in our 1D case. For example, the square lattice limits the accessible $\ket{\Lambda}$ configurations from the degenerate manifold due to the occupation constraint and increased connectivity. The system cannot facilitate having a superposition of $\ket{L}$ and $\ket{R}$ in the form $\ket{\mathcal{D}}$ in each dimer site since this leads to configurations with high-energy fluctuations. Considering the square lattice as stacked 1D chains, $\Omega_{\mu w}$ term gives rise to configurations of the form $\ket{\nobarfrac{L}{L}}$ or $\ket{\nobarfrac{R}{R}}$ in the neighboring rows where the occupation constraint is violated. This leads to spontaneous symmetry breaking with Neel (checkerboard) order with a square lattice configuration with alternating rows of $\ket{\tilde R}=(\ket{E}-\ket{R})/\sqrt{2}$ and $\ket{\tilde L}=(\ket{E}-\ket{L})/\sqrt{2}$. 

\begin{figure}[!t]
		\includegraphics[width=1\columnwidth]{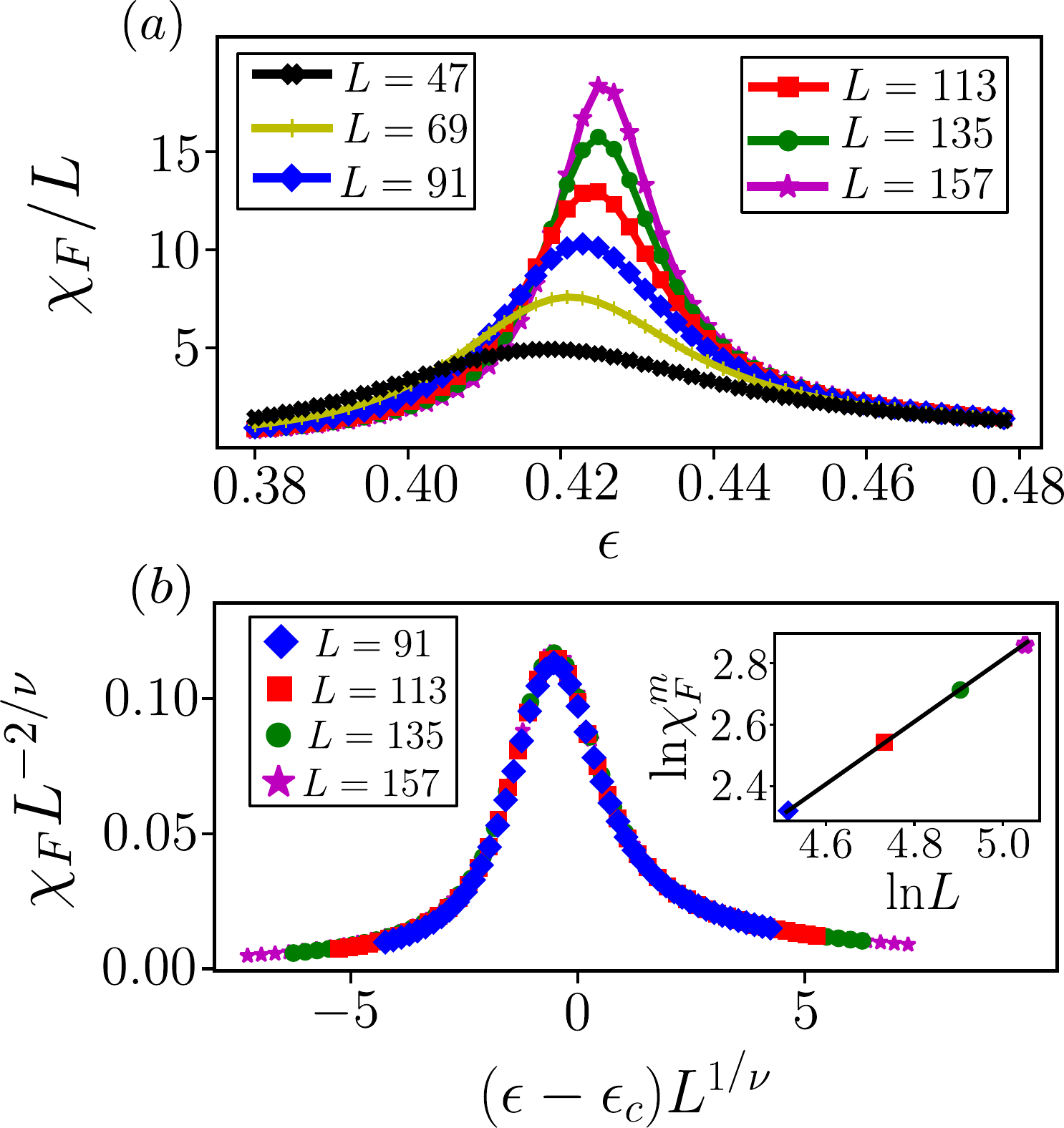}
		\caption{(a) The fidelity susceptibility per site $\chi_{F}/L$
exhibits a peak around the critical point. The peak becomes more pronounced as the system size $L$ gets larger. (b) Rescaled $\chi_F$ displays a data collapse for the critical point $\epsilon_c = 0.429$ and the critical exponent $\nu=1.011$ for different system sizes. Fidelity susceptibility at pseudocritical points $\chi_F^m$ as a function of system size in logarithmic scale is shown in the inset.} 
		\label{fig:FS}
\end{figure} 

In the regime $\epsilon \sim 0.5 - 1 $, both fluctuation terms are comparable in magnitude. This results in a crystalline phase with long-range order. In Fig.~\ref{fig:ord_par}(a), finite values of $O_{\text{CDW}}$ imply that the unit cells double and the translational symmetry is spontaneously broken. This is also corroborated by the behavior of the bipartite von Neumann entanglement entropy $S_{\text{vN}}\equiv -\Tr(\rho_r\ln{\rho_r})$ of the ground state as a function of $\epsilon$, where $\rho_r$ is the reduced density matrix of half of the chain. As seen in Fig.~\ref{fig:ord_par}(b), $S_{\text{vN}}$ makes a peak near the quantum critical point (QCP) and drops in the ordered phase. This vanishing of $S_{\text{vN}}$ around $\epsilon \sim 0.5$ [Fig.~\ref{fig:ord_par}(b)] implies that the ordered phase becomes an exact product state. As $\epsilon \to 1$, the ordered state becomes entangled yet still stays close to a product state form. CDW character is also reflected in the alternating density oscillations which imply that the bosons occupy every alternating site on the chain as shown in Fig.~\ref{fig:ord_par}(c). Differing from conventional CDW phases, the unit cell is given by $\ket{\tilde L}$ or $\ket{\tilde R}$ instead of $\ket{L}$ or $\ket{R}$ as given in Fig.~\ref{fig:ord_par}(d). We reveal the ordering mechanism by introducing both local and non-local fluctuations to the ordered system. This can be inferred from $\hat H_{\text{dim}} (\ket{E}-\ket{L})/\sqrt{2}=(\Omega_{\mu w} - t)\ket{R}- \Omega_{\mu w} (\ket{E}-\ket{L})/\sqrt{2}$, where local fluctuations couple $\ket{E}$ to $\ket{R}$ due to equal favoring of $\ket{L}$ and $\ket{R}$ whereas non-local fluctuations couple $\ket{L}$ to $\ket{R}$ due to $U(1)$ symmetry. Therefore, the system can be further constrained to prefer left dimers by tuning $t$ to lower the coupling to $\ket{R}$. This also explains why having $t<0$ in Eq.~(\ref{eqn:PXP}) does not help restore the long-range order but rather promotes the disordered state \cite{SM}. Therefore, having non-local fluctuations
mimics the increased connectivity of 2D by lowering the coupling of $\ket{\tilde L}$ ($\ket{\tilde R}$) to $\ket{R}$ $(\ket{L})$. This is also corroborated by the ansatz overlap $|\braket{\Psi|\Psi_{\text{ED}}}| \sim 0.99$ with variational parameters $u = 0.92, v = -0.92,w = 0.01$ \cite{SM}, where vanishing $w$ indicates that $\ket{L}$ is favored. Therefore, the ansatz leads to a product state of the form $\bigotimes_i \ket{\tilde L}_i$. The essential feature required to restore order in 1D is the necessity of combined local and non-local fluctuations. This emphasizes a crucial aspect of our setup, distinguishing it from constrained models in conventional Rydberg simulators. 

\begin{figure}[!t]
		\includegraphics[width=1\columnwidth]{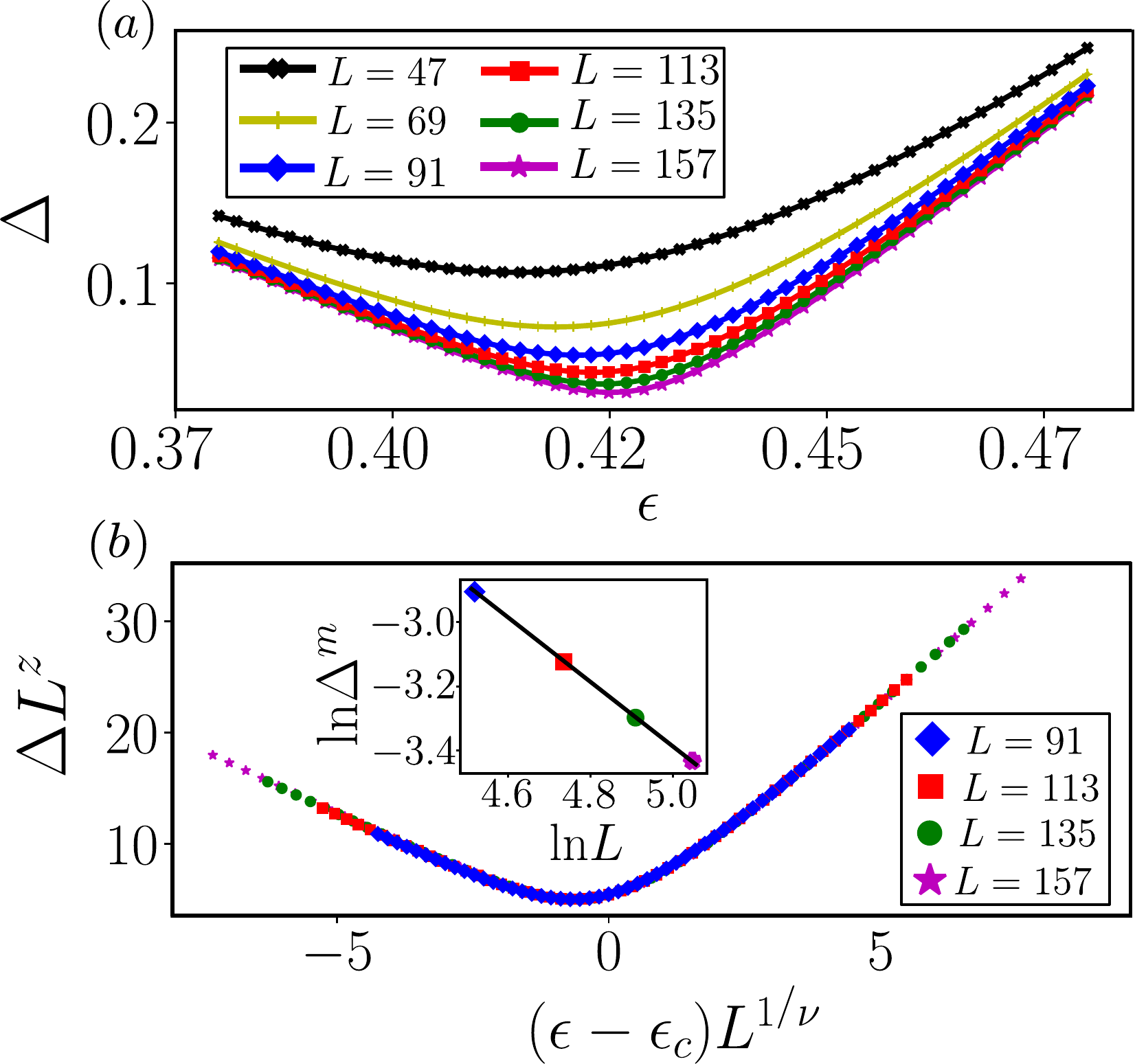}
		\caption{(a) $\Delta$ forms a dip in the vicinity of the critical point. As the system size $L$ gets larger, $\Delta$ approaches zero, thus, making the dip sharper. (b) Rescaled $\Delta$ exhibits a data collapse for the critical point $\epsilon_c = 0.429$, correlation critical exponent $\nu=1.011$, and the dynamic critical exponent $z=1.001$ for various system sizes. The inset depicts the neutral gap at pseudocritical points $\Delta^m$ as a function of the system size in logarithmic scale.} 
		\label{fig:NGAP}
\end{figure} 

We examine the nature of the QPT by first performing finite-size analysis of the fidelity susceptibility $\chi_{F}  = \lim\limits_{\mathrm{d} \epsilon \to 0} 2(1-\braket{\phi(\epsilon)|\phi(\epsilon+\mathrm{d} \epsilon)})/(\mathrm{d} \epsilon)^2 $\cite{Fid_sus_defin_2007, Fid_sus_defin_2008, Fid_Sus_Ryd_2022}, where $\ket{\phi(\epsilon)}$ is the ground state of $\hat H(\epsilon)$ in Eq.~(\ref{eqn:PXP}). $\chi_{F}$ probes the QPT by detecting the abrupt change in the ground-state character as $\epsilon$ varies. This can be seen in Fig.~\ref{fig:FS}(a) where the $\chi_{F}/L$ makes a peak around the QCP with the peak becoming more pronounced as the system size $L$ increases. This divergent behavior is captured in the inset of Fig.~\ref{fig:FS}(b) by analyzing the power-law behavior of the fidelity susceptibility given by $\chi_{F}^{m}(\epsilon_m, L) \propto L^{2/\nu}$ \cite{Fid_sus_defin_2007,Fid_suc_scal_2010} at pseudocritical points $\epsilon_m$, where $\nu$ is the correlation critical exponent and is extracted as $\nu=1.011$ by the slope of the fitted line. To locate the QCP, the finite-size scaling law for $\chi_{F}$ in the form $\chi_{F}(\epsilon, L)=L^{2/\nu}\mathcal{F}_{F} [L^{1/\nu}(\epsilon-\epsilon_c)]$ is analyzed, where $\epsilon_c$ is the QCP and $\mathcal{F}_{F}$ is an unknown scaling function \cite{Fid_suc_scal_2010, fid_sus_scaling_2008}. Fig.~\ref{fig:FS}(b) displays the behavior of the scaling function $\mathcal{F}_F$ for various system sizes by using rescaled variables $\chi_F L^{-2/\nu}$ and $(\epsilon-\epsilon_c)L^{1/\nu}$. A good data collapse is achieved by substituting the previously obtained exponent $\nu=1.011$ and tuning the $\epsilon_c=0.429$ as shown in Fig.~\ref{fig:FS}(b). A similar analysis of the neutral gap $\Delta = E_1 - E_0 $ for the scaling forms $\Delta^m(\epsilon_m,L)\propto L^{-z}$ and $\Delta(\epsilon, L)=L^{-z}\mathcal{F}_{\Delta} [L^{1/\nu}(\epsilon-\epsilon_c)]$ is performed \cite{det_scaling_func_2018}. Here, $E_0$ and $E_1$ are the ground and first excited state energy of $\hat H$ in Eq.~(\ref{eqn:PXP}). The dynamical critical exponent is determined as $z=1.001$ as shown in Fig.~\ref{fig:NGAP}(a,b). We conclude that the order-by-disorder transition belongs to the (1+1)D Ising universality class with $\nu=1$ and $z=1$. The key difference from earlier results is that combining local and non-local fluctuations drives the transition in the disordered-to-ordered direction where the long-range order is restored. 

\textit{Conclusion and outlook.\textemdash}Quantum fluctuations give rise to a wide range of phenomena in frustrated systems. This work promotes studying fluctuations in frustration-free systems using strongly interacting Rydberg atoms. Using vdW and dipolar interactions, a constrained model is derived, which differs from previous models \cite{PXP_FSS, PXP_REF1} by putting local and non-local fluctuations on an equal footing. Long-range order in 1D is established through mimicking high dimensionality where the increased connectivity is replaced by the synergistic effects of local and non-local fluctuations. This illustrates the combined effects of local and non-local fluctuations since they do not promote order individually. We developed a theoretical description of the model that offers a variational ansatz and explains the order formation. The ordered phase can in principle be verified experimentally for a system of tens of atoms a few microns apart from each other with $\{\ket{r=60S},\ket{r^{\prime}=61P}\}$ and $\Omega_{\mu w}$ of $1-5$ MHz \cite{zeybek_2023}. For future works, it would be interesting to see the physics arising from combining local and non-local fluctuations in higher dimensional lattices with occupation constraints.

\begin{acknowledgments}
		This work is funded by the Cluster of Excellence
		``CUI: Advanced Imaging of Matter'' of the Deutsche Forschungsgemeinschaft (DFG) - EXC 2056 - Project ID 390715994. This work is funded by the German Federal Ministry of Education and Research within the funding program ``quantum technologies - from basic research to market" under contract 13N16138.
\end{acknowledgments}

\bibliographystyle{apsrev4-1}
\bibliography{ref.bib}

\clearpage

\onecolumngrid
\newpage

\setcounter{equation}{0}            % reset equation counter
\setcounter{section}{0}    % reset section counter
\setcounter{figure}{0}    % reset section counter
\renewcommand\thesection{\arabic{section}}    % puts letters as section numbering
\renewcommand\thesubsection{\arabic{subsection}}    % puts letters as section numbering
\renewcommand{\thetable}{S\arabic{table}}
\renewcommand{\theequation}{S\arabic{equation}}
\renewcommand{\thefigure}{S\arabic{figure}}
\setcounter{secnumdepth}{2}
\setcounter{page}{1}

\begin{center}
	{\large Supplemental Material: \\ 
		\titleinfo}
\end{center}
\thispagestyle{empty}

\section{Derivation of the constrained model}

In this section, the constrained model given in Eq.~(\ref{eqn:PXP}) is derived starting from the Rydberg Hamiltonian given in Eq.~(\ref{eqn:EBHM_Ryd}) as,

\begin{equation}
    \hat H_{\text{Ryd}} = \sum_{i<j} t_{ij}( \hat b^{\dagger}_i \hat b_j +\text{h.c.}) + \sum_{i<j}V_{ij} \hat n_i \hat n_j   -  \Delta_{\mu w}\sum_{i}\hat n_i +\Omega_{\mu w}\sum_{i}(\hat b^{\dagger}_i+ \hat b_i),
\end{equation}
where $\hat b^{\dagger}_i$($\hat b_i$) is the bosonic creation (annihilation) operator at site $i$ with $(\hat b^{\dagger}_i)^2=0$, and $\hat n_i=\hat b^{\dagger}_i \hat b_i$ is the site density operator. Long-range hopping and density interaction strengths are denoted as $V_{ij}$ and $t_{ij}$ with $1/|i-j|^6$ and $1/|i-j|^3$ scaling, respectively. $\Omega_{\mu w}$ and $\Delta_{\mu w}$ stand for the microwave Rabi frequency and detuning. We continue working in the case of dominant nearest-neighbor processes where longer-range interaction and hopping terms are neglected ($t_{i,i+k}, V_{i,i+k} = 0, k>1$) with vanishing detuning ($\Delta_{\mu w}=0$). The resulting Hamiltonian is written as, 

\begin{equation}
    \hat H =  \underbrace{V\sum_i \hat n_i \hat n_{i+1}}_{\hat H_0} + \underbrace{t\sum_i (\hat b^{\dagger}_i \hat b_{i+1} +\text{h.c.}) +\Omega_{\mu w}\sum_{i}(\hat b^{\dagger}_i+ \hat b_i)}_{\hat H_p},
\end{equation}
where $V$ and $t$ denote the strengths of nearest-neighbor density interactions and hoppings, respectively. We are interested in the regime of strong interactions ($V \gg \Omega_{\mu w},t$) where the laser coupling and hopping terms are treated perturbatively. In this way, the first term is the non-perturbing $\hat H_0$ whose eigenstates are known exactly and can be grouped into eigenspaces $\mathcal{E}_\alpha, \mathcal{E}_\beta, \dots$ with different energies $E_{\alpha}$,$E_{\beta},\dots$ determined by the number of nearest-neighbor bosons. Therefore, a fixed eigenspace $\alpha$ consists of states with a definite number $N_{\bullet \bullet}$ of nearest-neighbor bosons. For example, $\mathcal{E}_{\alpha=0}$, $\alpha \equiv N_{\bullet \bullet}$, denotes the manifold of states with $N_{\bullet \bullet}=0$ nearest-neighbor boson occupations with energy $E_0 = 0$. As mentioned in the main text, the eigenspace $\alpha = 0$ is of great interest since it forms the required extensive classical ground state manifold. The combined second and third terms denote the perturbing $\hat H_p$ that couples states residing in different/same manifolds. We are interested in the effective Hamiltonian acting only within the $\alpha = 0$ manifold. Therefore, the effects of the perturbation are incorporated only within $\alpha = 0$ and high-energy processes causing inter-manifold couplings are integrated out. The effective Hamiltonian $\hat H_{\text{eff}}$ is defined up to second order in $\hat H_{\text{p}}$ by the non-zero matrix elements between any two states $\ket{k,\alpha}$ and $\ket{l,\alpha}$ with energy $E_{k\alpha}$ and $E_{l\alpha}$ in the same eigenspace $\alpha$ as \cite{QPT_Book_Sachdev_2011, SWT_Cohen, SWT_REV_2011},

\begin{equation}
    \braket{k,\alpha | \hat H_{\text{eff}} | l,\alpha} = E_{k\alpha}\delta_{kl}+ \braket{k,\alpha | \hat H_{\text{p}} | l,\alpha} + \frac{1}{2}\sum_{m, \beta \neq \alpha} \braket{k,\alpha | \hat H_{\text{p}} | m,\beta}\braket{m,\beta | \hat H_{\text{p}} | l,\alpha}\Big( \frac{1}{E_{k\alpha}-E_{m\beta}}+\frac{1}{E_{l\alpha}-E_{m\beta}}\Big).
\end{equation}

We are interested in the expression up to the first order, thus, the third term above is neglected. In the operator form, the above expression for $\alpha=0$ becomes,

\begin{equation} \label{eq:Heff}
    \hat H_{\text{eff}} = \hat H_0 \mathcal{\hat P}_{0} + \mathcal{\hat P}_{0}  \hat H_p \mathcal{\hat P}_{0},
\end{equation}
where $\mathcal{\hat P}_{0} = \prod_{\expval{ij}} (\mathbb{1}-\hat Q_i \hat Q_j)$ with $\hat Q_{i}=\ket{\bullet}_i\bra{\bullet}$ is the projection operator to the $\alpha = 0$ manifold of states without nearest-neighbor $\expval{ij}$ boson occupations. The first term above vanishes since $\hat H_0$ essentially counts the number of nearest-neighbor boson pairs, which is zero in the $\alpha = 0$ manifold. It is useful to decompose the perturbing term $\hat H_p $ into block-off-diagonal and block-diagonal forms with respect to the eigenspaces $\alpha$ of $\hat H_0$. In this way, the block-diagonal terms that encode intra-manifold couplings would survive the projector $\mathcal{\hat P}_{0}$, and the operator form of the second term in $\alpha$ is obtained. The decomposition is written as,

\begin{equation}
    \hat H_p = \Omega_{\mu w}(\hat \Omega_0 + \hat\Omega_1 + \hat\Omega_2 + \hat\Omega_{-1} + \hat\Omega_{-2})+ t(\hat T_0 + \hat T_1 + \hat T_{-1}),
\end{equation}
where each term above is given as follows,

\begin{equation}
	\hat \Omega_1 = \sum_i (\hat Q_{i-1} \hat b_i^{\dagger} \hat P_{i+1} + \hat P_{i-1} \hat b_i^{\dagger} \hat Q_{i+1}), \quad \hat\Omega_{-1} = \hat\Omega_1^{\dagger},
\end{equation}
where $\hat P_{i}=\ket{\circ}_i\bra{\circ}$. Here $\hat \Omega_1$ ($\hat \Omega_{-1}$) creates (annihilates) a boson at a given site $i$ if the site $i$ is surrounded by a boson on one of the sides. This process increases/decreases the number $N_{\bullet \bullet}$ in a given state by one, thereby, coupling manifolds of states differing from each other by a single pair of nearest-neighbor boson occupation, i.e., states in the eigenspace $\mathcal{E}_{\alpha=N}$ and $\mathcal{E}_{\alpha=N+1}$ are coupled.

\begin{equation}
	\hat \Omega_2 =  \sum_i \hat Q_{i-1} \hat b_i^{\dagger} \hat Q_{i+1}, \quad \hat\Omega_{-2} = \hat\Omega_2^{\dagger},
\end{equation}
Here $\hat \Omega_2$ ($\hat \Omega_{-2}$) creates (annihilates) a boson at a given site $i$ if the site $i$ is surrounded by bosons on both sides. This results in adding/removing the number $N_{\bullet \bullet}$ in a given state by two, thereby, coupling states in the eigenspace $\mathcal{E}_{\alpha=N}$ and $\mathcal{E}_{\alpha=N+2}$.

\begin{equation}\label{eq:OM0}
	\hat \Omega_0 = \sum_i \hat P_{i-1} (\hat b_i^{\dagger}+\hat b_i) \hat P_{i+1},
\end{equation}
Here $\hat \Omega_0$ creates and annihilates a boson at a given site $i$ if the site $i$ is surrounded by empty sites. Hence, $\hat \Omega_0$ does not add/remove nearest-neighbor boson pairs upon acting on states within a given manifold $\alpha$. Therefore, it only encodes intra-manifold couplings.

\begin{equation}
	\hat T_1 = \sum_{\expval{i,j}} (\hat Q_{i-1} \hat b_i^{\dagger} \hat b_j \hat P_{j+1} + \hat Q_{j+1}\hat b_j^{\dagger} \hat b_i \hat P_{i-1}), \quad \hat T_{-1} = \hat T_1^{\dagger}
\end{equation}
Here $\hat T_1$ represents the hopping of bosons between nearest-neighbor sites $i,j$ if the sites are surrounded only by a single boson on one of the sides. This operation ends up adding a pair of nearest-neighbor bosons.  Therefore it couples the manifold of states with $\mathcal{E}_{\alpha=N}$ and $\mathcal{E}_{\alpha=N+1}$. Analogously, $\hat T_{-1}$ removes a pair of nearest-neighbor bosons.
\begin{equation}\label{eq:T0}
		\hat T_0 = \sum_{\expval{i,j}} (\hat Q_{i-1}b_i^{\dagger} \hat b_j \hat Q_{j+1}+ \hat P_{i-1}b_i^{\dagger} \hat b_j \hat P_{j+1} + \text{h.c.}).
\end{equation}
Here $\hat T_0$ represents the hopping of bosons between nearest-neighbor sites $i,j$ if the sites $i,j$ involved in the particle exchange are surrounded by bosons (first term) and empty sites (second term). $\hat T_0$ does not add or remove a pair of nearest-neighbor bosons upon acting on states within a given manifold $\alpha$ thereby encoding only intra-manifold couplings. 

Since we are interested only in intra-manifold couplings, the only relevant terms for Eq.~(\ref{eq:Heff}) come from Eq.~(\ref{eq:OM0}) and (\ref{eq:T0}) giving,
\begin{equation}
    \hat H_{\text{eff}} = \mathcal{\hat P}_{0}[\Omega_{\mu w}\sum_i \hat P_{i-1}(\hat b^\dagger_i +\hat b_i)  \hat P_{i+1} + t\sum_{i}(\hat Q_{i-1}b_i^{\dagger} \hat b_j \hat Q_{j+1}+ \hat P_{i-1}b_i^{\dagger} \hat b_j \hat P_{j+1} + \text{h.c.})]\mathcal{\hat P}_{0},
\end{equation}
where the projectors $\mathcal{\hat P}_{0} = \prod_{\expval{ij}} (\mathbb{1}-\hat Q_i \hat Q_j)$ get rid of the $\hat Q_{i-1}(b_i^{\dagger} \hat b_j +\text{h.c.}) \hat Q_{j+1}$ in $\hat T_0$ since it violates the occupation constraint where no two bosons are allowed to be adjacent. The above equation is written in the $\alpha=0$ manifold of states obeying the occupation constraint as follows,
\begin{equation}
    \hat H_{\text{eff}} = \Omega_{\mu w}\sum_i \hat P_{i-1}(\hat b^\dagger_i +\hat b_i)  \hat P_{i+1} + t\sum_{i}\hat P_{i-1}(\hat b^\dagger_i \hat b_{i+1} + \text{h.c.})\hat P_{i+2}.
\end{equation}

\section{Mapping to Interacting Dimers}
In this section, the constrained model given by Eq. (\ref{eqn:PXP}) in the main text is expressed by interacting dimers where two sites are combined into one. The resulting Hamiltonian acting on the dimer states with the appropriate operators is derived. Variational many-body ansatz inspired by the dimer mapping is proposed. 

After dimerizing the lattice as shown in Fig.~\ref{fig:sett}(b) in the main text, the configurations that represent the state of a dimer are given by $\{\ket{\circ \circ}, \ket{\circ \bullet},\ket{\bullet \circ}\}$. In this way, the state $c_1\ket{\circ \circ}+c_2\ket{\circ \bullet}+c_3\ket{\bullet \circ}$ represents the most general dimer state. However, this construction does not respect the occupation constraint since it implicitly assumes that dimers are independent of each other. This can be seen from the fact that the state of a given dimer does not depend on the nearby dimers. To remedy this problem and impose the occupation constraint, creating a dimer at a site $j$ of the form $\ket{\circ \bullet}$ is accompanied by attaching a fermion to the right bond $(j,j+1)$. Similarly, creating a dimer at a site $j$ of the form $\ket{\bullet \circ}$ is accompanied by attaching a fermion to the left bond $(j-1,j)$ as depicted in Fig.~(\ref{fig:sett}) in the main article. Such auxiliary fermions have been employed in other works as well \cite{ChengI, Cheng_II, ChengIII}. This procedure encodes empty, left, and right-dimer states which make up the dimer degree of freedom as ${\ket{E},\ket{R},\ket{L}}$. Single-site and dimer degrees of freedom are equivalent since the latter corresponds to distinguishing even- and odd-site boson occupations in the former. This could be seen by considering the left-dimer as odd-site boson occupied states and the right-dimer as even-site boson occupied states. By using the dimer degrees of freedom, we define the following operators,

\begin{align}\label{eq:LROPS}
    \hat d^{\dagger}_{L,j}\hat f^{\dagger}_{j-1,j} &=\ket{\bullet \circ}_j \bra{\circ \circ} \otimes \ket{1}_{j-1,j} \bra{0} \notag\\
    \hat d^{\dagger}_{R,j}\hat f^{\dagger}_{j,j+1} &=\ket{\circ \bullet}_j \bra{\circ \circ} \otimes \ket{1}_{j,j+1} \bra{0},
\end{align}
where $\hat d^{\dagger}_{L,j}\hat f^{\dagger}_{j-1,j}$ creates a left-dimer at site $j$
with a fermion between site $j$ and $j-1$ ($\ket{1}_{j,j-1}$) and $\hat d^{\dagger}_{R,j}\hat f^{\dagger}_{j,j+1}$ creates a right-dimer at site $j$ with a fermion between site $j$ and $j+1$ ($\ket{1}_{j,j+1}$). This can be seen in the following,

\begin{align}\label{eq:L}
    \hat d^{\dagger}_{L,j}\hat f^{\dagger}_{j-1,j}\ket{E} = \hat d^{\dagger}_{L,j}\hat f^{\dagger}_{j-1,j} \ket{0}_{j-1,j}\otimes\ket{\circ \circ}_{j} \otimes \ket{0}_{j,j+1} &= \ket{\bullet \circ}_j \bra{\circ \circ} \otimes \ket{1}_{j-1,j} \bra{0}[\ket{0}_{j-1,j}\otimes\ket{\circ \circ}_{j} \otimes \ket{0}_{j,j+1}] \notag \\ &= \ket{1}_{j-1,j}\otimes\ket{\bullet \circ}_{j}\otimes \ket{0}_{j,j+1} = \ket{L}_j,
\end{align}

\begin{align}\label{eq:R}
    \hat d^{\dagger}_{R,j}\hat f^{\dagger}_{j,j+1}\ket{E} = \hat d^{\dagger}_{R,j}\hat f^{\dagger}_{j,j+1} \ket{0}_{j-1,j}\otimes\ket{\circ \circ}_{j} \otimes \ket{0}_{j,j+1} &= \ket{\circ \bullet}_j \bra{\circ \circ} \otimes \ket{1}_{j,j+1} \bra{0}[\ket{0}_{j-1,j}\otimes\ket{\circ \circ}_{j} \otimes \ket{0}_{j,j+1}] \notag \\ &= \ket{0}_{j-1,j}\otimes\ket{\circ \bullet}_{j}\otimes \ket{1}_{j,j+1} = \ket{R}_j,
\end{align}
where $\ket{L}$ and $\ket{R}$ are shown above in Eq.~(\ref{eq:L}) and (\ref{eq:R}), respectively. The Pauli exclusion of auxiliary fermions imposes occupation constraint. For example, a state of the form $\ket{\cdots R_k L_{k+1} \cdots }$ is discarded since $ \ket{\cdots R_k L_{k+1} \cdots } = \hat d_{R,k}^{\dagger} \hat f_{k,k+1}^{\dagger}  \hat d_{L,k+1}^{\dagger} f_{k,k+1}^{\dagger}\ket{\cdots E_k E_{k+1} \cdots}$, where $f_{k,k+1}^{\dagger}f_{k,k+1}^{\dagger}$ violates Pauli exclusion principle. By using the operators defined in Eq.~(\ref{eq:LROPS}), we will encode the local- and non-local processes in Eq.~(\ref{eqn:PXP}) in the main article using the dimer degree of freedom. The local fluctuation is provided by the Rabi $\Omega_{\mu w}$ term and is expressed by the following,
\begin{align}
    \Omega_{\mu w} \sum_j ( \hat d^{\dagger}_{L,j}\hat f^{\dagger}_{j-1,j} + \hat d^{\dagger}_{R,j}\hat f^{\dagger}_{j,j+1} + \text{h.c.}),
\end{align}
where the creation and annihilation of left/right dimer states are given. Specifically, the first term corresponds to creation/annihilation processes on odd-numbered sites through the mapping $d^{\dagger}_{L,j} \hat f^{\dagger}_{j-1,j} \to \hat{P}_{i-1}\hat b^{\dagger}_{i}\hat P_{i+1}$ with $i=2j-1$. With this identification, the effect of the right projector $\hat{P}_{i+1}$ is a built-in property due to $\ket{\bullet \circ}$ in $\ket{L}$ and the auxiliary fermion on the left link encodes the left projector $\hat{P}_{i-1}$. Analogously, the second term corresponds to creation/annihilation processes on even-numbered sites through the mapping $\hat d^{\dagger}_{R,j} \hat f^{\dagger}_{j,j+1} \to \hat{P}_{i-1}\hat b^{\dagger}_{i}\hat P_{i+1}$ with $i=2j$. The non-local fluctuation is provided by the hopping $t$ term and is expressed by the following,

\begin{equation}
    t\sum_j (\hat d^{\dagger}_{L,j}\hat f^{\dagger}_{j-1,j}\hat d_{R,j}\hat f_{j,j+1}+ \text{h.c.} ) + t\sum_{\expval{j,k}}(\hat d^{\dagger}_{R,j}\hat f^{\dagger}_{j,j+1}\hat d_{L,k}\hat f_{k,k-1}+ \text{h.c.}),
\end{equation}
where the first term corresponds to the flipping between $\ket{L}_j$ and $\ket{R}_j$ locally at dimer-site $j$. Since this happens locally, it encodes intra-dimer hopping processes over even and odd sites in terms of single-site states. For example, let us consider the case of dimer-site $j=2$ which includes single-sites $i=3,4$. Therefore, the flipping of $\ket{R}_2$ to $\ket{L}_2$ gives rise to boson hopping from $i=4$ to $i=3$. This could be seen by the explicit substitution of the single-site operators into the dimer hopping expression. $\hat d^{\dagger}_{L,2}\hat f^{\dagger}_{1,2}\hat d_{R,2}\hat f_{2,3} = (\hat{P}_{2}\hat b^{\dagger}_{3}\hat P_{4}) (\hat{P}_{3}\hat b_{4}\hat P_{5}) = \hat{P}_{2}\hat b^{\dagger}_{3} \hat b_{4}\hat P_{5}$, since operators residing on different sites commute and $\hat b^{\dagger}_3 \hat P_3 = \hat b^{\dagger}_3$, $\hat P_4 \hat b_4 = \hat b_4$. The second term corresponds to the exchange between $\ket{L}_j$ and $\ket{R}_k$ at nearest-neighbor dimer-sites $\expval{j,k}$. Since this involves two dimer-sites, it encodes inter-dimer hopping processes over even and odd sites in terms of single-site states. For example, consider the exchange of $\ket{L}_3$ with $\ket{R}_2$ which is expressed as $\hat d^{\dagger}_{R,2}\hat f^{\dagger}_{2,3}\hat d_{L,3}\hat f_{2,3} = (\hat{P}_{3}\hat b^{\dagger}_{4}\hat P_{5}) (\hat{P}_{4}\hat b_{5}\hat P_{6}) = \hat{P}_{3}\hat b^{\dagger}_{4}b_{5}\hat P_{6}$ since operators residing on different sites commute and $\hat b^{\dagger}_4 \hat P_4 = \hat b^{\dagger}_4$, $\hat P_5 \hat b_5 = \hat b_5$. Therefore, both local and non-local processes in the single-site description are generated with the dimer degree of freedom. Motivated by the dimer mapping we propose the following ansatz,

\begin{figure}[t!]
		\includegraphics[width=\columnwidth]{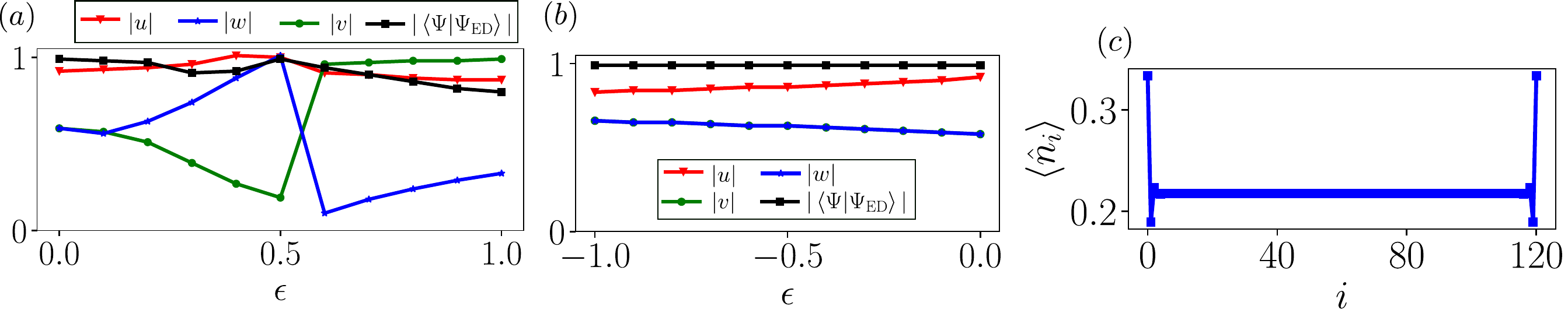}
		\caption{(a,b) Overlap of the $\ket{\Psi_\text{ED}}$ with the ansatz $\ket{\Psi}$ with optimal variational parameters $(u,v,w)$ is shown. A system of size $L=20$ with periodic boundary conditions is considered for the exact diagonalization. (c) DMRG simulation for the expectation value of the density operator $\hat n_i$ for a system size $L=121$ with $\epsilon=-0.5$ is shown. In contrast to Fig.~\ref{fig:ord_par}(c) in the main article, the disordered state with homogeneous density is obtained.} 
		\label{fig:SManstz}
\end{figure}  

\begin{equation}
    \ket{\Psi} = \frac{1}{\mathcal{N}}\prod_j^{N} (u+ v \hat d^{\dagger}_{L,j}\hat f^{\dagger}_{j-1,j}+ w \hat d^{\dagger}_{R,j}\hat f^{\dagger}_{j,j+1})\ket{E \cdots E},
\end{equation}
where $\ket{E \cdots E}$ is the many-body state of empty dimers with no fermions on the links. $N$ denotes the number of dimer sites. Variational parameters are $u,v,w$ and $\mathcal{N}$ corresponds to the normalization constant. The state $\ket{\Psi}$ amounts to superposing all many-body dimer configurations $\ket{\Lambda}$ that obey the occupation constraint. To illustrate this, the ansatz above is expanded as follows,
\begin{align}
    \ket{\Psi} &= (u+v \hat d_{L,1}^{\dagger} + w \hat d_{R,1}^{\dagger} \hat f_{1,2}^{\dagger})(u+v \hat d_{L,2}^{\dagger} f_{1,2}^{\dagger} + w \hat d_{R,2}^{\dagger}f_{2,3}^{\dagger}) \cdots \ket{E \cdots E} \notag \\
    &= (u^2 + uv\hat d_{L,2}^{\dagger} f_{1,2}^{\dagger}+ uw\hat d_{R,2}^{\dagger} + uv \hat d_{L,1}^{\dagger} + v^2\hat d_{L,1}^{\dagger} d_{L,2}^{\dagger} f_{1,2}^{\dagger}+ vw d_{L,1}^{\dagger} \hat d_{R,2}^{\dagger}f_{2,3}^{\dagger} + uw  \hat d_{R,1}^{\dagger} \hat f_{1,2}^{\dagger} \notag \\ &+ \cancel{ vw  \hat d_{R,1}^{\dagger} \hat f_{1,2}^{\dagger}  \hat d_{L,2}^{\dagger} f_{1,2}^{\dagger}} + w^2 \hat d_{R,1}^{\dagger} \hat f_{1,2}^{\dagger} \hat d_{R,2}^{\dagger}f_{2,3}^{\dagger}) (u+v \hat d_{L,3}^{\dagger} f_{2,3}^{\dagger} + w \hat d_{R,3}^{\dagger}f_{3,4}^{\dagger})   \cdots \ket{E \cdots E} \notag \\
    &= (u^N+ u^{N-1}v\hat d_{L,1}^{\dagger} + u^{N-1}w\hat d_{R,1}^{\dagger}\hat f_{1,2}^{\dagger} \cdots + v^N \hat d_{L,1}^{\dagger}\hat d_{L,2}^{\dagger} f_{1,2}^{\dagger} \cdots \hat d_{L,N}^{\dagger} f_{N-1,N}^{\dagger}  + w^N \hat d_{R,1}^{\dagger}\hat f_{1,2}^{\dagger} \hat d_{R,2}^{\dagger}\hat f_{2,3}^{\dagger} \cdots \hat d_{R,N}^{\dagger}) \ket{E \cdots E} \notag \\
    &= u^N \ket{E \cdots E} +  u^{N-1}v\ket{LE \cdots E}+ u^{N-1}w\ket{RE \cdots E} + \cdots + v^N\ket{L \cdots L} + w^N\ket{R \cdots R} \notag \\
    &= \sum_{\Lambda} u^{N^{\Lambda}_{E}}  v^{N^{\Lambda}_{L}} w^{N^{\Lambda}_{R}} \ket{\Lambda}
\end{align}
where terms that create different many-body dimer configurations $\ket{\Lambda}$  are given inside the brackets. In this way, all the $\ket{\Lambda}$  configurations obeying the occupation constraint are generated. For example, the configuration that violates the constraint is crossed out due to having $\hat f_{1,2}^{\dagger}\hat f_{1,2}^{\dagger}$. The final expression is written in terms of $\ket{\Lambda}$ as given in Eq.~(\ref{eq:ansatz}) in the main article. To test the performance of the ansatz, we compute the optimal parameters $(u,v,w)$ by minimizing the variational energy $\braket{\Psi | \hat H_{\text{dim}} | \Psi}$, where periodic boundary conditions applied. We then calculate the overlap of the optimal ansatz with the exact ground state of $\hat H_{\text{dim}}$ obtained by exact diagonalization with periodic boundary conditions. In the disordered regime ($\epsilon \to 0$), the overlap $|\braket{\Psi|\Psi_{\text{ED}}}| \sim 0.99$ with $v=w$ is achieved as shown in Fig.~\ref{fig:SManstz}(a). This implies that the superposition exhibits equal favoring of $\ket{L}$ and $\ket{R}$ in the many-body configurations $\ket{\Lambda}$. Motivated by this observation, we further find out that the optimal superposition can be written as $\ket{\Psi_{\text{dis}}}=\bigotimes_i \ket{\mathcal{D}}_i$ with $\ket{\mathcal{D}}_i=[(\ket{E}_i/\sqrt{2})-(\ket{L}_i+\ket{R}_i) /2]$, where $i$ denotes a single dimer-site. Numerical calculations of the overlaps show that $|\braket{\Psi_{\text{dis}}|\Psi_{\text{ED}}}| \sim 0.99$ and $|\braket{\Psi_{\text{dis}}|\Psi}| \sim 0.99$ with $v=w$. As the system enters the ordered regime, equal favoring of $\ket{L}$ and $\ket{R}$ as given in $\ket{\mathcal{D}}$ is no longer the case. This is reflected in the ansatz with $v \neq w$. Large overlaps in the ordered regime $\epsilon \in [0.5,1.0]$ are obtained as shown in Fig.~\ref{fig:SManstz}(a) (the lowest being $|\braket{\Psi|\Psi_{\text{ED}}}| \sim 0.88$). As mentioned in the main article, when $t<0$ long-range order is not restored and the disordered state with homogeneous density [Fig.~\ref{fig:SManstz}(c)] is promoted. This can be seen from Fig.~\ref{fig:SManstz}(b) where $|\braket{\Psi|\Psi_{\text{ED}}}| \sim 0.99$ with $v=w$.  

\end{document}